\documentclass[preprint,aps]{revtex4}
\usepackage{graphicx}
\usepackage{dcolumn}
\usepackage{bm}
\newcommand{\be}{\begin{equation}}
\newcommand{\ee}{\end{equation}}

\begin{document}


\title{A~new multi-center approach to the 
exchange-correlation interactions in ab initio tight-binding methods}

\author{Pavel Jel\'{\i}nek$^{1,2}$, Hao Wang$^{3}$, James P.
Lewis$^{3}$, Otto F. Sankey$^{4}$ and Jos\'e Ortega$^{1}$}

\affiliation{$^{1}$ Departamento de F\'{\i}sica Te\'orica de la
Materia Condensada, Universidad Aut\'onoma de Madrid, E-28049
Spain}

\affiliation{$^2$ Institute of Physics,
Academy of Sciences of the  Czech Republic,
Cukrovarnick\'a 10, 1862 53,Prague, Czech Republic}

\affiliation{$^{3}$ Department of Physics and Astronomy,
Brigham Young University, Provo, Utah 84602-4658, USA}

\affiliation{$^{4}$ Department of Physics and Astronomy, Arizona
State University, Tempe, Arizona 85287-1504, USA}

\date{\today}
\begin{abstract}
A~new approximate method to calculate exchange-correlation contributions 
in the framework of first-principles tight-binding molecular dynamics 
methods has been developed. In the proposed scheme on-site (off-site)
exchange-correlation matrix elements are expressed as 
a~one-center (two-center) term plus a~{\it correction} due to the
rest of the atoms. The one-center (two-center) term is evaluated
directly, while the {\it correction} is calculated
using a~variation of the Sankey-Niklewski \cite{Sankey89} approach 
generalized for arbitrary atomic-like basis sets. The proposed scheme for 
exchange-correlation part permits the accurate and computationally efficient 
calculation of corresponding tight-binding matrices and atomic forces for 
complex systems.  We calculate bulk properties of selected 
transition (W,Pd), noble (Au) or simple (Al) 
metals, a~semiconductor (Si) and the transition metal oxide Ti$O_2$ 
with the new method to demonstrate its flexibility and good accuracy.  

\end{abstract}
\maketitle
\section{Introduction}
The application of first-principles simulation techniques is
becoming a research tool of increasing importance in materials
science, condensed matter physics and chemistry, and molecular
physics and chemistry. Most of these techniques are based on
Density Functional Theory (DFT) \cite{Hohenberg64} which creates an
important simplification of the many-body quantum mechanical
problem. Typically, DFT calculations are performed within the Kohn-Sham
approach \cite{Kohn65} using the Local Density Approximation
(LDA)\cite{Kohn65} or a Generalized Gradient Approximation
(GGA)\cite{Becke}. These total energy quantum mechanical methods can
be used to calculate forces on atoms, and thus
perform first-principles molecular dynamics (MD) simulations.
Such simulations have been very successful in the description of
a variety of properties of different materials. But, in spite of the 
important simplifications introduced by DFT and related approximations 
({\it e.g.} LDA,GGA), they still require a huge amount of computational 
resources. This problem has severely limited the range of applications 
of these simulation techniques to situations with small numbers of atoms 
($\sim$ 100-200) in the unit-cell, and short simulation times.

Due to the computational limitations, first-principles simulation techniques
have been mainly directed to the study of the energetics and
electronic structure of diverse materials, surfaces and molecules.
Typically, a good guess for the atomic structure is
obtained before the calculation, and the first-principles method
is then used to refine the geometry, obtain the electronic
structure, and to compare the total energy of a few competing
structures. These methods, however, have been very rarely applied
to elucidate complex atomic structures, that require the {\it
exploration} of an~extensive phase space of possibilities, when no 
{\it a~priori} answer, or approximate good guess, is already available.
More importantly, the application of first-principles methods to
investigate complex kinetic processes in materials ({\it e.g.} the
atomic motion of atoms on a surface, kinetic pathways, molecular
reactions, etc) is still very limited, due to the computational
resources required for these calculations.

It is clear that the usefulness of first-principles
simulation techniques can be greatly extended if appropriate
approximations are made, with the purpose of increasing
the computational efficiency, with as little loss of accuracy as possible
\cite{Sankey89,Porezag,Siesta,Horsfield97,Horsfield00,OO-LCAO}. 
This idea has prompted the development of first-principles tight-binding
molecular dynamics (TBMD) methods
\cite{Sankey89,Horsfield97,Ortega98a,Horsfield00}, whose main
characteristics are: (1)  {\it a real-space} technique ({\it i.e.}
no need for super-cells or grids), (2) {\it optimized atomic-like
orbitals} \cite{Eschrig78,Sankey89,Horsfield97,Junquera01} as basis set, and
 (3)  efficient, two-dimensional, tabulation-interpolation schemes
\cite{Sankey89,Horsfield97} to obtain the effective TB Hamiltonian
matrix elements as well as their derivatives to obtain the forces.

The main advantage of such techniques is computational
efficiency which makes them ideal first-principles exploratory
tools. The use of first-principles
TBMD methods as a exploratory tool can be complemented with
more accurate calculations, if necessary; once stimulating results or
ideas are obtained, final results can be refined by
performing more accurate and time-consuming calculations
(plane-waves DFT - {\it e.g.}\cite{Ortega00} - or even many-body -
{\it e.g.} \cite{Ortega98b} - calculations).

In this paper we report on new developments for the treatment of
exchange-correlation contributions in first-principles TBMD
methods, and their implementation in the {\sc Fireball} code
\cite{Lewis01,Demkov95,Sankey89}. The approximations used so far
to handle these contributions are analyzed in Section II. In
Section III we present our new approach to calculate the
exchange-correlation contributions, and in Section IV we present some 
results for several materials to illustrate the performance of 
the new approach.

\section{Ab initio tight binding: Fireball}
{\sc Fireball} \cite{Lewis01,Demkov95,Sankey89} is a
first-principles TBMD simulation technique based on a~self-consistent 
version of the Harris-Foulkes \cite{Harris85,Foulkes89} functional. 
The energy functional is written as:
\begin{equation}
E_{TOT} [ \rho(\vec r) ] = \sum_n\varepsilon_n-E_{ee}[\rho(\vec r)]
+E_{xc}[\rho(\vec r)] -
\int {\rho(\vec r)V_{xc}[\rho(\vec r)] d^3 r}
+ E_{ion-ion},
\label{Harris}
\end{equation}
where $\rho(\vec r)$ is the {\it input density}, which will be allowed to
vary, and will be determined selfconsistently.
The first term is a sum over occupied eigenstates,
$\varepsilon_n$, of the effective one-electron Hamiltonian,
\begin{eqnarray}
\Biggl (-{1\over {2}} \nabla^2 + V [\rho] \Biggr)\psi_n =
\varepsilon_n \psi_n; \label{Heff}
\end{eqnarray}
the potential $V$ is the sum of the ionic potential, $v_{ion}(\vec
r)$, (typically represented by a pseudopotential), a Hartree
potential, and an exchange-correlation potential $V_{xc}$
\begin{eqnarray}
V [\rho] = v_{ion}(\vec r)
 + \int {\rho(\vec {r^{\prime}}) d^3 {r^{\prime}}
\over |\vec r - \vec r^{\prime}| } + V_{xc} [\rho(\vec r)].
\label{Veff}
\end{eqnarray}
In equation (\ref{Harris}) $E_{ee}$ is an average electron-electron energy,
\begin{eqnarray}
E_{ee} [\rho] =
 \frac{1}{2} \int \int \frac{\rho (\vec r) \rho (\vec{r} \, ')}{| \vec r -
\vec{
r} \, ' |} d\vec r d\vec{r} \, ',
\end{eqnarray}
$E_{ion-ion}$, is the ion-ion interaction energy
\begin{eqnarray}
E_{ion-ion} =
\frac{1}{2} \sum_{i,j} \frac{Z_{i} Z_{j} }{| \vec
R_{i}
 - \vec R_{j} |}
\end{eqnarray}
($Z_{i}$ is the nuclear or pseudopotential charge of atom $i$ at
position $\vec R_{i}$),
and $E_{xc}[\rho]$ is the exchange-correlation energy.
First-principles MD simulations can be performed once the forces
\begin{eqnarray}
\vec F_{i} = - \frac{\partial {E_{TOT}}
}{\partial \vec
R_{i} }
\label{forces}
\end{eqnarray}
on each atom $i$ are evaluated.

The efficiency of calculations based on the Harris functional is associated
 with the possibility to choose $\rho({\vec r})$ in the above equations as a sum of
atomic-like densities, $\rho_i({\vec r})$ :
\begin{equation}
\rho({\vec r}) = \sum_i \rho_i({\vec r}). \label{rhosumi}
\end{equation}
In the {\sc Fireball} method, confined atomic-like orbitals
are used as a basis set for the determination of the occupied
eigenvalues and eigenvectors of the one-electron Hamiltonian, Eq.
(\ref{Heff}). The fireball orbitals, introduced by Sankey and
Niklewski (SN) \cite{Sankey89}, are obtained by solving the atomic
problem with the boundary condition that the atomic orbitals
vanish outside and at a predetermined radius $r_c$ where $\psi(\vec r)|_{r=r_c}= 0$ (see Fig. 1).
An important advantage of the fireball basis set is that the Hamiltonian
(Eqn. (\ref{Heff})) and the overlap matrix elements are quite sparse for large
systems. The electron density $\rho({\vec r})$ is written in
terms of the fireball orbitals $\phi_{ilm}({\vec
r})\equiv\phi_{\mu}({\vec r})$ ($i$ is the atomic site, $l$
represents the atomic subshell - {\it e.g.} $3s,3p,3d,$ etc., 
and $m$ is the magnetic quantum number)
\begin{eqnarray}
\rho({\vec r})=\sum_\mu q_\mu |\phi_\mu (\vec r )|^2.
\label{rho}
\end{eqnarray}
In this way four-center integrals are not required for the calculation of the
Hartree terms, and all the two- and three center interactions are
tabulated beforehand and placed in interpolation data-tables which are 
no larger than two-dimensional \cite{Sankey89}. Hamiltonian matrix elements 
are evaluated by looking up the necessary information from the data-tables.

In practice, the atomic densities $\rho_i$
\begin{eqnarray}
\rho_i({\vec r})=\sum_{lm} q_{ilm} |\phi_{ilm} (\vec r )|^2.
\label{rho_ilm}
\end{eqnarray}
are approximated to be spherically symmetric around each atomic
site $i$
({\it i.e.} $q_{ilm} = q_{ilm'}$). Self-consistency is achieved by
imposing that the output orbital charges {$q_\mu^{out}$} (obtained
from the occupied eigenvectors $\psi_n$ of equation (\ref{Heff}))
and input orbital charges $q_{\mu}$  coincide (see ref. \cite{Demkov95} 
and \cite{Lewis02} for further details).


The remaining difficulty is the efficient calculation of exchange-correlation
interactions within a~first-principles TB scheme. One possibility is to
use non-standard DFT and introduce the exchange-correlation energy and 
potential as a~function of the orbital occupancies \cite{OO-LCAO,FJ,Pou}.
In this paper, however, we opt for the more traditional approach in which
exchange-correlation contributions are calculated as a~functional of the
electron density $\rho (\vec r)$. Within this line, two different methods
have been previously proposed for the practical calculation of 
exchange-correlation terms, using data-tables similar to those for the
Hartree contributions. These two methods are:   



\subsection{Sankey-Niklewski approximation}
The basic idea introduced by SN \cite{Sankey89}
is to write down the non-linear in $\rho ({\vec r})$ exchange-correlation 
matrix elements
in terms of matrix elements of $\rho ({\vec r}) $ . These later matrix
elements are easily tabulated in data-tables no larger than two-dimensional,
similar to those required for the Hartree terms.

Consider the matrix elements $< \phi_\mu | V_{xc} [\rho] | \phi_\nu>$ of 
the exchange-correlation potential. For each matrix element 
$< \phi_\mu | V_{xc} [\rho] | \phi_\nu>$, expand $V_{xc} [ \rho ]$ 
in a Taylor's series
\begin{equation}
V_{xc} [ \rho ] \simeq V_{xc} [ \bar{\rho}_{\mu\nu} ]
+ V'_{xc} [ \bar{\rho}_{\mu\nu} ] (
\rho -  \bar{\rho}_{\mu\nu} ) + ...
\end{equation}
around an appropriate ``average density" $\bar{\rho}_{\mu\nu}$:
\begin{equation}
\bar{\rho}_{\mu\nu} =
{ < \phi_\mu | \rho | \phi_\nu >
\over
< \phi_\mu | \phi_\nu > }
\end{equation}
With this choice of
$\bar{\rho}_{\mu\nu}$ the second term in the expansion for
$< \phi_\mu | V_{xc} [\rho] | \phi_\nu >$ is zero, and the next term is
minimized \cite{Sankey89}. This yields
\begin{equation}
< \phi_\mu | V_{xc} [\rho] | \phi_\nu >
\equiv <\mu | V_{xc} [\rho] |\nu > \simeq
V_{xc} [ \bar{\rho}_{\mu\nu} ] < \mu | \nu >
\end{equation}
SN \cite{Sankey89} realized that corrections to this
approximation are required for the case $ <\phi_\mu | \phi_\nu >
\equiv <\mu | \nu
> = 0$, and they devised a scheme to add those corrections specifically for 
a minimal $sp^3$ basis set.

 The first step towards improved an~improved exchange-correlation 
contributions in our TBMD scheme 
(see Sec. III) will be, precisely, to extend their ideas for a general 
atomic-like basis set. Also, it will be shown below that
 the SN average density approximation is not accurate for the
exchange-correlation-energy on-site terms
$<\mu | \epsilon_{xc} |\mu >$ for atoms with a significant valence electron 
density such as transition metals (e.g. Au, Ag, Pd, etc.); in Sec. III 
we will present a new scheme that corrects this problem.

 \subsection{Horsfield approximation}

An alternative approach to deal with exchange-correlation terms
within a first-principles TBMD method was proposed by Horsfield
\cite{Horsfield97}, who introduced a many-center expansion based on Eqn.
(\ref{rhosumi}). In this approach we can distinguish two cases 
\cite{Horsfield97} ($i_\mu$ is the atomic site corresponding 
to orbital $\mu$ and $i_\nu$ corresponds to orbital $\nu$),

(a) $i_\mu = i_\nu \equiv i $
\begin{eqnarray}
<\mu | V_{xc} [\rho] |\nu > \simeq
<\mu |V_{xc}[\rho_i]|\nu > + \sum_{j \neq i} <\mu |(V_{xc}[\rho_i +
\rho_j] - V_{xc}[\rho_i])|\nu >,
\label{HXC1}
\end{eqnarray}

(b) $(i_\mu \equiv i) \neq (i_\nu \equiv j) $
\begin{eqnarray}
<\mu | V_{xc} [\rho] |\nu > =
<\mu |V_{xc}[\rho_i + \rho_j]|\nu > +
\sum_{k \neq i,j} <\mu | ( V_{xc}[\rho_i + \rho_j +\rho_k] -
V_{xc}[\rho_i + \rho_j] |\nu >.
\label{HXC2}
\end{eqnarray}

Although practical experience has shown that this is an accurate
approach in many cases, the on-site terms ( case (a) ) are not
always well-approximated by the above equation, and additional
numerical integrals are necessary for those
term \cite{Horsfield97,Horsfield00}. Another shortcoming of this
approach is the fact that most of the computational time required
to create the data-tables within this approximation is spent in the
calculation of the exchange-correlation terms, reducing the
computational efficiency.

\section{A new exchange-correlation scheme for Ab initio tight-binding }
We now present a new approximation to calculate
exchange-correlation contributions in a first-principles TBMD
method. This new approximation is composed of two main parts. First, 
we generalize the SN approximation for an arbitrary
atomic-like basis set. Second, we propose an improved approximation which 
combines the best features of both the SN and Horsfield
approximations.

To generalize the SN approach beyond $\rm sp^3$ basis sets, we define average
densities $\bar{\rho}_{\mu\nu}$ using auxiliary, spherically
symmetric orbitals $\varphi_{il}$ (defined below). The use of these spherically symmetric 
orbitals solves problems associated with the zero overlap $< \mu | \nu
> = 0$ cases; moreover, this approach for calculating exchange-correlation 
terms is consistent with the
spherical approximation used in calculating  Hartree terms.


We define spherically symmetric orbitals $\varphi_{il}\equiv
\varphi_{\mu}$ for each atomic subshell $(i,l)$, corresponding to
atomic-like orbitals $\phi_{ilm}$ as follows. First, we consider the 
atomic-like orbitals

\begin{equation}
\phi_{ilm} = \omega_{il}(r) Y_{lm} (\Omega)
\end{equation}
where $\omega_{il}(r)$ is the radial part of $\phi_{ilm}$ and $Y_{lm}
(\Omega)$ the spherical harmonic associated with the angular part.
Next, we define the spherically symmetric orbitals by
\begin{equation}
\varphi_{il} = \tilde{\omega}_{il}(r) Y_{00} (\Omega)
\end{equation}
where $\tilde{\omega}$ is the positive root of $(\omega)^2$ (see
Fig. 1); $\tilde{\omega}$ is defined this way in order to avoid
spurious cancellations in the importance sampling calculation of 
$\bar{\rho}_{\mu\nu}$ (Eq. \ref{av-rho}) which occur with certain intra-atomic 
cases (e.g. two different s-orbitals on the same atom).

\begin{figure}[ht]
  \centering
\vspace*{1.1cm}
\includegraphics[width=7.5cm]{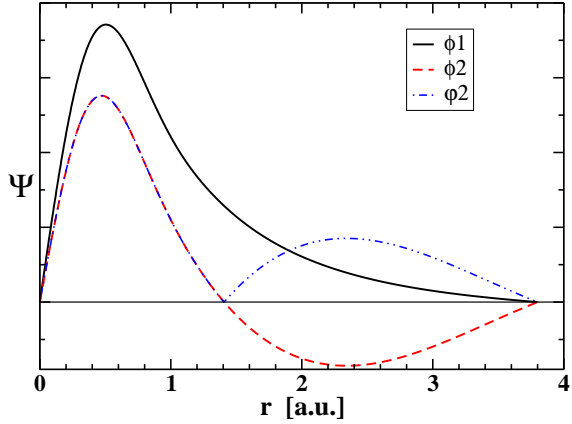}
  \caption{Two radial functions corresponding to Fireball $\phi_1$ and $\
	phi_2$, and auxiliary $\varphi_2$ orbital. The auxiliary 
	orbital $\varphi_1$ (not shown) coincides with $\phi_1$.}
  \label{basis}
\end{figure}


With these auxiliary orbitals we now define average
densities for each matrix element $(\mu,\nu)$ as
\begin{equation}
\bar{\rho}_{\mu\nu} = { < \varphi_{\mu} | \rho | \varphi_{\nu} >
\over
< \varphi_{\mu} | \varphi_{\nu} > }
\label{av-rho}
\end{equation}
This new definition for the average densities $\bar{\rho}_{\mu\nu}$
(using the auxiliary orbitals $\varphi$ instead of the atomic
orbitals $\phi$) solves all problems related to zero overlap
($< \phi_\mu | \phi_\nu > = 0$), since now $< \varphi_{\mu} |
\varphi_{\nu} > \neq 0$; moreover, the use of auxiliary
orbitals represents an improvement in the ``importance
sampling" calculation of $\bar{\rho}_{\mu\nu}$ for the non-zero
overlap cases. Regions of positive overlap are no longer
``artificially" canceled by regions of negative overlap: both
positive and negative overlap regions add-up now in this new definition
of $\bar{\rho}_{\mu\nu}$.

Note that since the orbitals $\varphi_{il}$ are spherically symmetric,
the same value of $\bar{\rho}_{\mu\nu}$ is obtained for all the matrix
elements ($\mu,\nu$) associated with a given pair of atomic 
subshells ($i,l;i',l'$).
Also with our new definition of $\bar{\rho}_{\mu\nu}$, we have 
in general $ <\mu |\rho | \nu> \neq \bar{\rho}_{\mu\nu} <\mu |
\nu>$, and we keep the next term in the Taylor's expansion for
$V_{xc} [ \rho ]$:
\begin{equation}
<\mu | V_{xc} [\rho] |\nu > \simeq
V_{xc} [ \bar{\rho}_{\mu\nu} ] < \mu | \nu >
+ V'_{xc} [ \bar{\rho}_{\mu\nu} ] \biggl( < \mu | \rho | \nu > -
\bar{\rho}_{\mu\nu} < \mu | \nu > \biggr)
\label{SNXC}
\end{equation}
We refer to this approximation as the generalized SN approach (GSN approach).

The average density approximation is not very accurate for
certain matrix elements, particularly $<\mu|\epsilon_{xc}|\mu>$. 
This is not a serious problem (as we demonstrate in Sec. IV) 
for the determination of the structural and/or electronic properties, 
since basically only  the absolute value of the total energy is 
affected ({\it i.e.} it represents a rigid shift in the total energy). 
Nevertheless, the next step in our new approximation is to correct 
this inaccuracy. For this purpose, we use the best features of the SN 
and the Horsfield approximations. As in the Horsfield scheme, 
we distinguish two cases: (a) on-site ($i_\mu = i_\nu $), and 
(b) off-site matrix elements.

(a) $i_\mu = i_\nu \equiv i $. As a first step in our approximation, 
we simply add and subtract a contribution associated with 
the atomic density
$\rho_i$ at site $i$, and write, {\it formally}, the matrix
element as a one-center contribution plus a {\it correction}, in
similarity with the Horsfield approach:
\begin{eqnarray}
<\mu | V_{xc} [\rho] |\nu > =
<\mu |V_{xc}[\rho_i]|\nu > + \biggl( <\mu |V_{xc}[\rho]|\nu >
- <\mu |V_{xc}[\rho_i]|\nu > \biggr).
\end{eqnarray}
The one-center (first) term is easily calculated and tabulated, and we
use the GSN  approach discussed above to evaluate the
{\it correction}:
\begin{eqnarray}
<\mu | V_{xc} [\rho] |\nu > & \simeq & <\mu |V_{xc}[\rho_i]|\nu >
\nonumber \\
& + & V_{xc} [ \bar{\rho}_{\mu\nu} ] < \mu | \nu > + V'_{xc} [
\bar{\rho}_{\mu\nu} ] \biggl( < \mu | \rho | \nu > -
\bar{\rho}_{\mu\nu} < \mu | \nu > \biggr)
\nonumber \\
& - & V_{xc}[{\bar{\rho}_i}]<\mu|\nu>
- V'_{xc}[\bar{\rho}_i]
\biggl( <\mu|\rho_i|\nu> - \bar{\rho_i} <\mu|\nu> \biggr)
\label{OLSXC1}
\end{eqnarray}
with
\begin{equation}
\bar{\rho}_i = { < \varphi_{\mu} | \rho_i | \varphi_{\nu} >
\over
< \varphi_{\mu} | \varphi_{\nu} > }
\end{equation}
(indices $\mu,\nu$ have been omitted in $\bar{\rho}_i$, for
clarity).

(b) $(i_\mu = i) \neq (i_\nu =j) $. Proceeding in a similar manner
as for the on-site matrix elements, we obtain for the off-site matrix elements:
\begin{eqnarray}
<\mu | V_{xc} [\rho] |\nu > & = &
<\mu |V_{xc}[\rho_i + \rho_j]|\nu > + \biggl( <\mu |V_{xc}[\rho]|\nu >
- <\mu |V_{xc}[\rho_i + \rho_j]|\nu > \biggr)
\\
& \simeq &
<\mu |V_{xc}[\rho_i + \rho_j]|\nu >
\nonumber
\\
& + & V_{xc} [ \bar{\rho}_{\mu\nu} ] < \mu | \nu >
+ V'_{xc} [ \bar{\rho}_{\mu\nu} ] \biggl( < \mu | \rho | \nu > -
\bar{\rho}_{\mu\nu} < \mu | \nu > \biggr)
\nonumber
\\
& - & V_{xc}[\bar{\rho}_{ij}]<\mu|\nu>
- V'_{xc}[\bar{\rho}_{ij}]
\biggl( <\mu|(\rho_i + \rho_j)|\nu> - \bar{\rho_{ij}} <\mu|\nu> \biggr)
\label{OLSXC2}
\end{eqnarray}
with
\begin{equation}
\bar{\rho}_{ij} = { < \varphi_{\mu} | (\rho_i + \rho_j) |
\varphi_{\nu} > \over < \varphi_{\mu} | \varphi_{\nu} > }
\end{equation}
(indices $\mu,\nu$ omitted for clarity). In Eqs. (\ref{OLSXC1}) and
(\ref{OLSXC2}) $\bar{\rho}_{\mu\nu}$, which includes all density contributions,
is defined using Eq. (\ref{av-rho}).

\section{Results}

In this section we present results illustrating the
performance of our new exchange-correlation scheme discussed in
section III.
Transition metals contain a significant valence electron
density (the $d$-electrons), mixed with a free-electron-like density 
(the $sp$-bands), and thus represent good test cases for the different 
exchange-correlation schemes. In this section we present results for some 
transition metals (Au, Pd, W), and the transition metal oxide $\rm TiO_2$; 
we also show some results for a~typical $sp^3$ metal (Al) and 
a~semiconductor (Si).

\begin{figure}[ht]
  \centering
\vspace*{1.0cm}
\includegraphics[width=7.5cm]{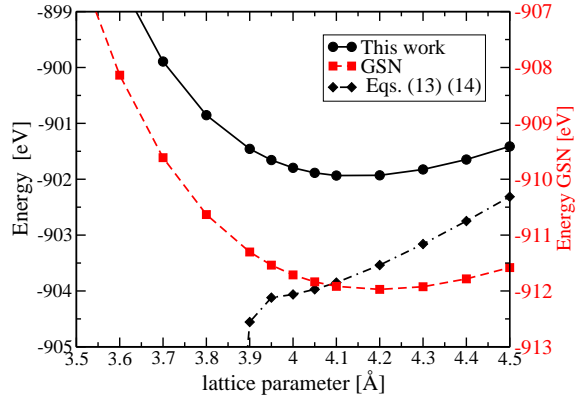}
  \caption{Total energy for bulk Au as a function of the lattice
parameter $a$ for different XC-approximations presented in this work. 
The energy scale on the right corresponds to the GSN approximation.
'This work' line shows the result obtained from Eqs. \ref{OLSXC1} 
and \ref{OLSXC2}. The dashed line is the result obtained using 
Eqs. \ref{HXC1} and \ref{HXC2}.  Basis set: $sp^3d^5$ fireball orbitals
with cut-off radii $R_c(s) = 4.6$ a.u., $R_c(p) = 5.2$ a.u. and
$R_c(d) = 4.1$ a.u.}
  \label{Etot-Au}
\end{figure}

As an example we plot energy vs. lattice constant for Au. 
Fig. \ref{Etot-Au} shows the total energy of bulk Au as a
function of the lattice parameter as calculated with the {\sc
Fireball} code using the new exchange-correlation scheme proposed
in section III ({\it i.e.} Eqs. (\ref{OLSXC1}) and
(\ref{OLSXC2})).
We used a $sp^3d^5$ basis set with fireball orbitals defined
by the following cut-off radii: $R_c(s) = 4.6$ a.u., $R_c(p) =
5.2$ a.u. and $R_c(d) = 4.1$ a.u. Also we show in the results for 
GSN (equation (\ref{SNXC})), and Horsfield
(obtained using Eqs. (\ref{HXC1}) and (\ref{HXC2}), {\it i.e.}
without the additional numerical integral for the on-site terms).
These results demonstrate how critical it is for the transition metals 
to have a good description of the on-site XC-contributions. 
Note the nearly rigid shift of the GSN result (scale on the right of
Fig. \ref{Etot-Au}) by $\sim$ 10 eV. As mentioned
in Secs. II and III, this is related to the inaccuracy of the
average-density approximation (Eq. \ref{SNXC}) for calculating
the on-site energy terms $<\mu | \epsilon_{xc} | \mu >$: Au
atoms contain a large electron density ( $\sim$ 10 $d$-electrons
plus 1 $s$-electron) in the valence band. For comparison, this shift 
is only $\sim 0.7$ eV in the case of bulk Al.

\begin{table}[htbp]
\caption{ Equilibrium lattice constants $a$ and bulk moduli $B$
for selected elements obtained using Eqs. \ref{OLSXC1} and \ref{OLSXC2} 
(This work), and Eq. \ref{SNXC}
(GSN) for the exchange-correlation LDA contributions.
We calculated these with $sp^3$ (Al,Si) or
$sp^3d^5$ (transition metals) basis sets of fireball orbitals with
cut-off radii $R_c$ (in a.u.) as indicated. Also shown
are PW-LDA and experimental values.} \label{tab:results}
\begin{ruledtabular}
\begin{tabular}{|c|c|c|c|c|c|c|c|c|c|c|}
        & & & \multicolumn{4}{c|}{ $a$ (\AA)} & \multicolumn{4}{c|}
{ $B (GPa)$} \\ \cline{4-11} Name  & Structure & Basis set   &
This work & GSN & PW-LDA & Expt. & This work & GSN & PW-LDA & Expt. \\
\hline
Au  & fcc & s4.6-p5.2-d4.1  & 4.14 & 4.21 & 4.06 & 4.07  & 210 & 197 &
205 & 171  \\
Pd  & fcc & s4.6-p5.0-d4.0  & 3.96 & 3.98 & 3.85 & 3.89  & 215 & 294 &
220 & 181  \\
W   & bcc & s4.7-p5.2-d4.5  & 3.18 & 3.17 & 3.14 & 3.17  & 347 & 320 &
333 & 310  \\
Si  & zbd & s4.8-p5.4  & 5.46 & 5.50 & 5.38 & 5.43  & 109 & 98  & 96 & 100  \\
Al  & fcc & s5.3-p5.7  & 4.04 & 4.09 & 3.97 & 4.05  & 93  & 85  & 84 & 76  \\
\end{tabular}
\end{ruledtabular}
\end{table}

Table I shows the calculated lattice parameter $a$ and Bulk
modulus $B_0$ (obtained using a Murnaghan equation of state EOS)
for Au, as well as for other transition metals (Pd, W), Al (a
typical free-electron-like metal) and for Si (a typical
semiconductor). These results have been obtained using either
minimal $sp^3$ basis sets (Al,Si) or $sp^3d^5$ basis sets
(transition metals). The experimental values \cite{exp-values},
and the plane-waves LDA values (PW-LDA), are also
presented in Table I. This Table shows that with the new approach
to introduce exchange-correlation contributions the experimental
lattice constants $a$ are reproduced within $ \sim 2$\% while the
bulk moduli are slightly overestimated by $\sim 15$\%. The
agreement is improved when comparing with the PW-LDA. Since the
accuracy of first-principles TBMD methods is mainly related with
the quality of the atomic-like basis set,
improvements on the results presented in Table I are to be
expected with a better choice for the basis set, either by
improving the $sp^3$ or $sp^3d^5$ orbitals {\it and/or} adding new
orbitals to the basis set ({\it e.g.} double basis sets, etc.)
\cite{Kenny00,Junquera01,Anglada02,Ozaki04}.

\begin{figure}[ht]
  \centering
\vspace*{0.1cm}
\includegraphics[width=7.5cm]{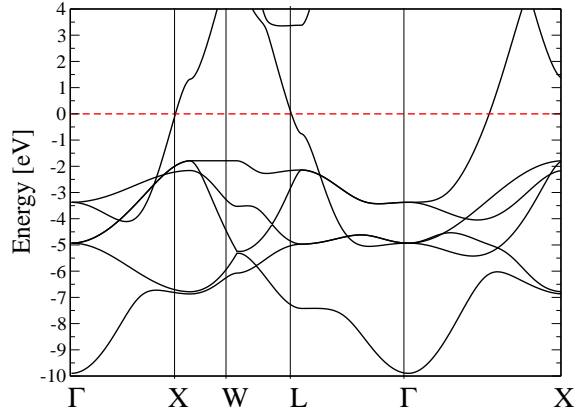}
  \caption{Band structure of Au fcc. LDA exchange-correlation terms are
  calculated using the new approximation discussed in Sec. III. 
  Basis set: $sp^3d^5$ fireball orbitals
  with cut-off radii $R_c(s) = 4.6$ a.u., $R_c(p) = 5.2$ a.u. and
 $R_c(d) = 4.1$ a.u..The dashed line represents the Fermi level.  }
  \label{Band-Au}
\end{figure}
\begin{figure}[ht]
  \centering
\vspace*{0.1cm}
\includegraphics[width=7.5cm]{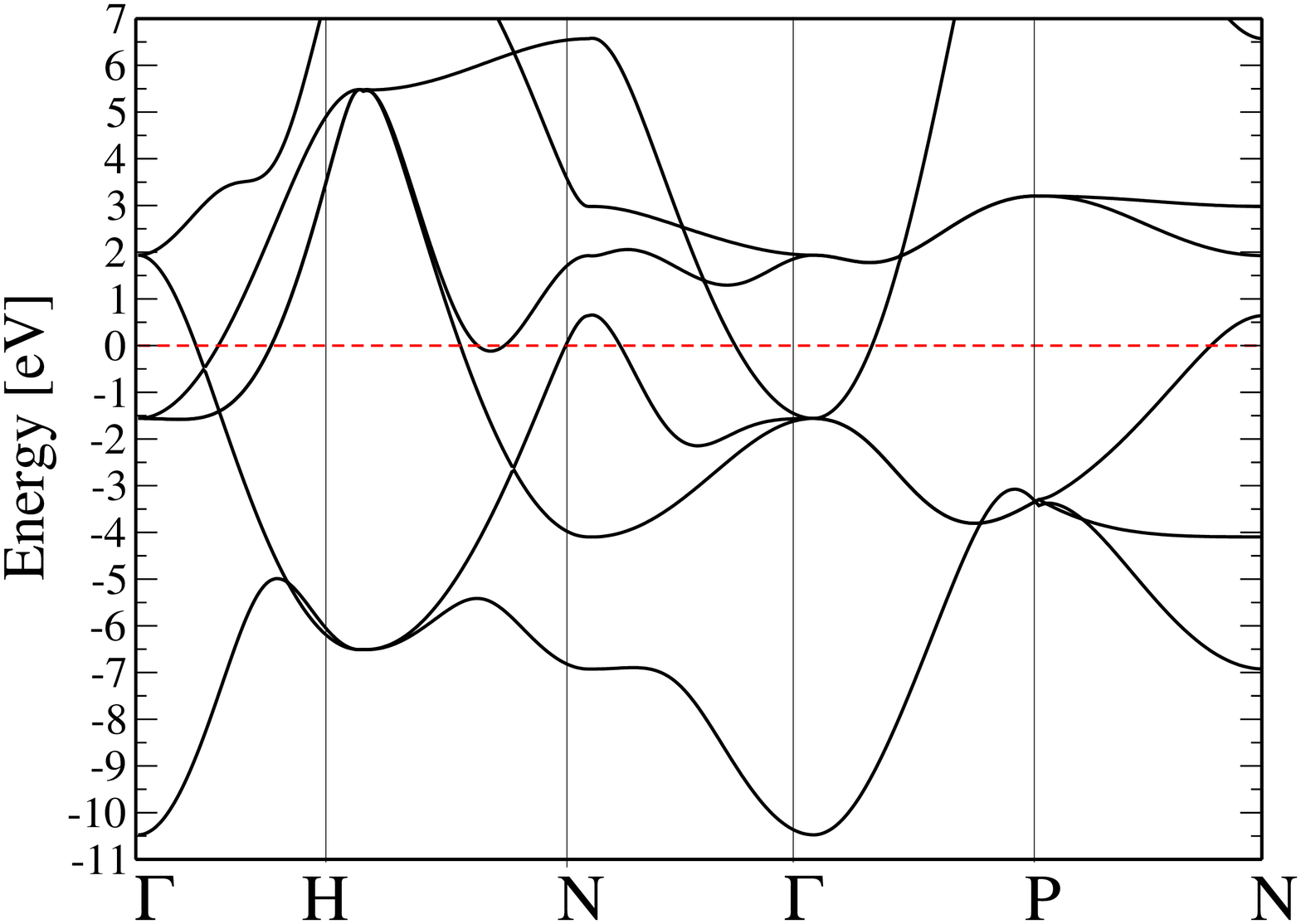}
  \caption{Band structure of W bcc (see also Fig. \ref{Band-Au}).
           Basis set: $sp^3d^5$ fireball orbitals with cut-off radii 
           $R_c(s) = 4.7$ a.u., $R_c(p) = 5.2$ a.u. and $R_c(d) = 4.5$ a.u..}
  \label{Band-W}
\end{figure}
\begin{figure}[ht]
  \centering
\vspace*{0.1cm}
\includegraphics[width=7.5cm]{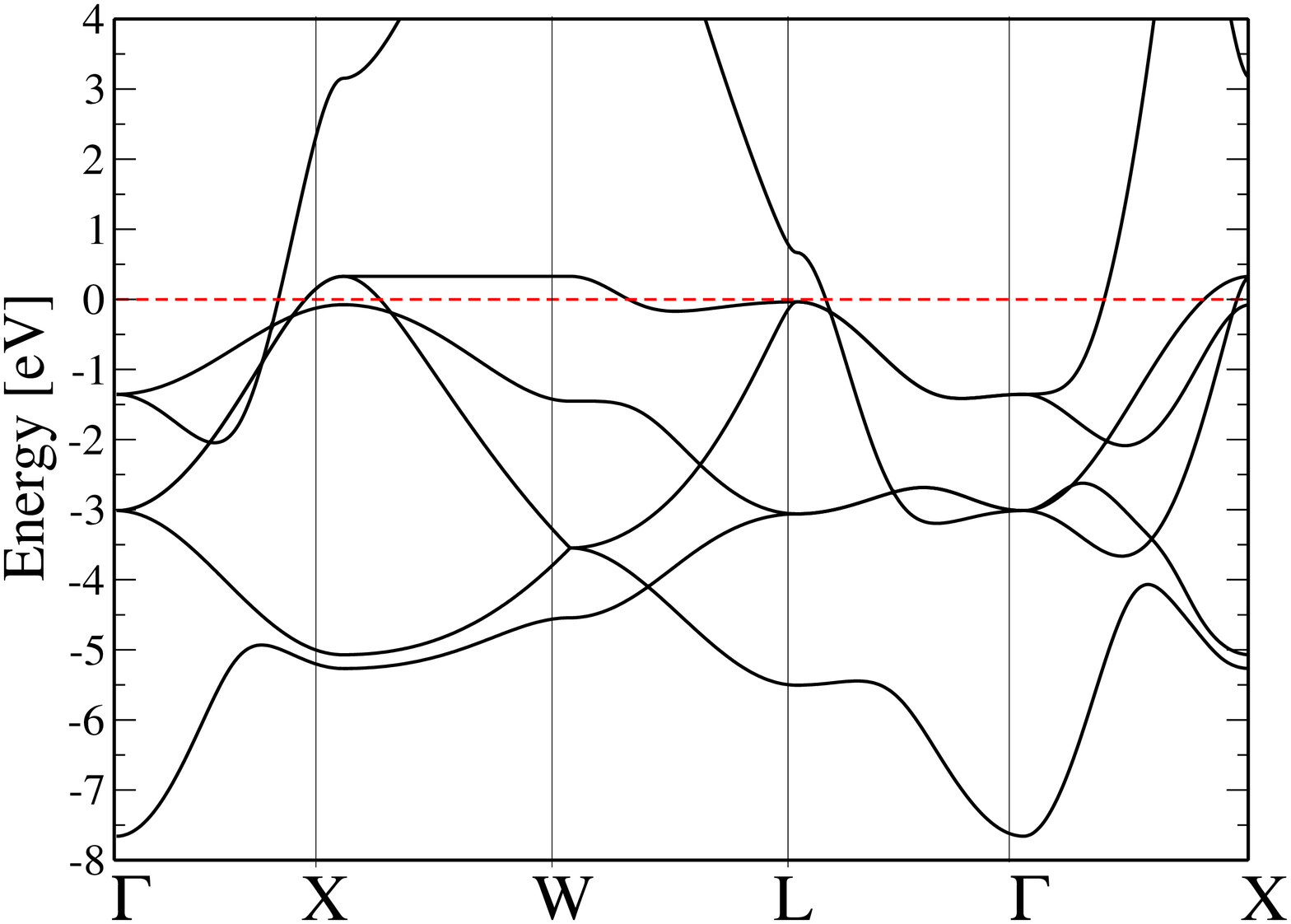}
  \caption{Band structure of Pd fcc (see also Fig. \ref{Band-Au}).
           Basis set: $sp^3d^5$ fireball orbitals with cut-off radii 
           $R_c(s) = 4.6$ a.u., $R_c(p) = 5.0$ a.u. and $R_c(d) = 4.0$ a.u..}
  \label{Band-Pd}
\end{figure}

Figures ~\ref{Band-Au},\ref{Band-W},\ref{Band-Pd} show the
band-structures for the transition metals Au, W and Pd. The
comparison with more accurate calculations ({\it e.g.} see
\cite{Papa}) shows that the band-structures are well-approximated
within the present first-principles TB-scheme. In this case, there
is practically no difference between the band-structures obtained
with the GSN or the new approach (equations (\ref{OLSXC1},\ref{OLSXC2})).



Tetragonal rutile structure TiO$_2$ belongs to the space group
$P4_{1}/mnm$, containing 6 atoms per unit cell. The structural
parameters for rutile structure TiO$_2$ have been determined to a
high degree of accuracy from the neutron diffraction experiments
performed by Burdett \emph{et al.} \cite{Burdett873639}. We have
calculated the structural parameters and electronic bandstructure
for TiO$_2$ in the rutile structure using the {\sc Fireball} code
and the exchange-correlation approach discussed in section III.
For these calculations we have used a $sp^3$ basis for oxygen with
cut-off radii $R_c(s) =3.6$ a.u. and $R_c(p)=4.1$ a.u., while for
Ti a basis set of $sp^3d^5$ orbitals was used, with cut-off radii
$R_c(s)=6.3$ a.u., $R_c(p)$=6.0 a.u. and $R_c(d)=5.7$ a.u. The
optimal structure is obtained by minimizing the total energy of
the rutile ($P4_{1}/mnm$) structures with respect to the lattice
parameters $a$, $c$, and internal parameter $u$. We perform this
minimization by a two-step procedure as outlined in
Ref.\cite{Chelikowsky921284}. Table \ref{cap:lattice comparison}
summarizes the comparison of our results to the experimentally
determined structural and elastic parameters in
TiO$_{\textrm{2}}$; results from other theoretical works are also
listed. This table shows that our results for the structural
properties of TiO$_2$ in the rutile structure are within $1$\% of
the experimental results of Burdett, \emph{et al.}
\cite{Burdett873639}. From the integrated EOS, we obtain a value
for the bulk modulus $B$ of 206 GPa which agrees well with the
experimental value of 211 GPa \cite{Arlt0014414}. In addition, our
results agree well with the calculated results of others
\cite{Chelikowsky921284, Mo9513023}.

\begin{table}[htpb]

\caption{\label{cap:lattice comparison}Theoretical results for
structural and elastic parameters for TiO$_{\textrm{2}}$ in the
rutile structure. Comparisons are made between our results and
experimental results for the volume  $V$, lattice parameters 
$a$, $c$, internal parameter  $u$, and bulk modulus  $B$; zero
subscript represents the experimental results \cite{Burdett873639}. }

\begin{center}\begin{tabular}{|c|c|c|c|c|c|c|}
\hline & $V/V_{0}$& $a/a_{0}$& $c/c_{0}$& $u/u_{0}$& $B$ (GPa)&
$B_{0}$ \cite{Arlt0014414} (GPa)\tabularnewline \hline \hline
Present work& 0.994& 0.997& 0.999& 0.994& 206& 211\tabularnewline
\hline Other calculation \cite{Chelikowsky921284}& 1.039& 1.013&
1.002& 1.001& 240 & \tabularnewline \hline Other calculation
\cite{Mo9513023} & 1.021& 0.999 & 1.002 & 0.998& 209 &
\tabularnewline \hline
\end{tabular}\end{center}
\end{table}

\begin{figure}[htpb]
  \centering
\vspace*{1.1cm}
\includegraphics[width=7.5cm]{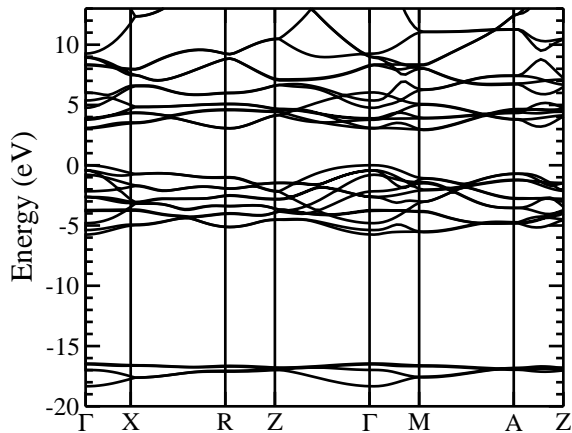}
  \caption{Band structures for TiO$_2$ in the rutile structure. The valence-band maximum is taken as the zero of
energy.}
  \label{Band-TiO2}
\end{figure}

Using our theoretically predicted equilibrium lattice parameters,
we have calculated the self-consistent electronic band structure
for rutile TiO$_2$ depicted in Fig. \ref{Band-TiO2} along the
high-symmetry directions of the irreducible Brillouin zone. Table
\ref{cap:band results} gives a summary of our results in
comparison to experiment and other calculations for the detailed
features of the band structure. The upper valence band is composed
of $O_{2p}$ orbitals and has a width of 5.75 eV. These results are
in agreement with the experimental values of 5.50 eV
\cite{Kowalczyk77161}. The lower $O_{2s}$ band is 1.89 eV wide.
Our results are consistent with other calculations
\cite{Chelikowsky921284,Mo9513023}. The calculated direct band gap
at $\Gamma$ of 3.05 eV is in agreement with the reported
experimental gap of 3.06 eV \cite{Pascual785606}. The LDA
generally underestimates the experimental band gap for insulators
and semiconductors, and the band gap obtained from $ab$ $initio$
plane-wave calculations for TiO$_2$ is $\sim$ 2.0 eV
\cite{Chelikowsky921284}. This underestimating effect of LDA is
compensated in our results because we use a local orbital basis
set. We also find an indirect band gap from $\Gamma$ to $M$ which
is smaller than the direct band gap by 0.13 eV. 

\begin{table}[htpb]
\caption{\label{cap:band results} Comparison of our present work
to the experimentally determined electronic properties for
TiO$_{\textrm{2}}$ in the rutile structure. Definition of listed
quantities are as follows: (1) E$_{\textrm{g}}$ - D is the direct
bandgap ($\Gamma$ to $\Gamma$), (2) E$_{\textrm{g}}$ - ID is the
indirect bandgap ($\Gamma$ to $M$), (3) E$_{\textrm{VB}}$ is the
upper valence bandwidth, and (4) E$_{\textrm{O2s}}$ is the Oxygen
$2s$ state bandwidth. }

\begin{center}\begin{tabular}{|c|c|c|c|c|}
\hline Structure& E$_{\textrm{g}}$ - D (eV)& E (eV)$_{\textrm{g}}$
- ID& E$_{\textrm{VB}}$ (eV)& E$_{\textrm{O2s}}$ (eV)
\tabularnewline \hline \hline Present & 3.05& 2.92& 5.75&
1.89\tabularnewline \hline Experiments & 3.06
\cite{Pascual785606}& & 5.50 \cite{Kowalczyk77161}&
\tabularnewline \hline Others& 1.78 \cite{Mo9513023}, 2.00
\cite{Chelikowsky921284}& 3.00 \cite{Mo9513023}, 2.00 \cite{Chelikowsky921284}
& 6.22 \cite{Mo9513023}, 5.7 \cite{Chelikowsky921284}& 1.94
\cite{Mo9513023}, 1.80 \cite{Chelikowsky921284}\tabularnewline
\hline
\end{tabular}\end{center}
\end{table}

\section{Summary}
In summary, we have presented a new approach to calculate
exchange-correlation contributions in first-principles TBMD
methods. After a brief presentation of the basic theoretical
foundations and practical motivation for these techniques, we have
discussed the different approximations ( SN \cite{Sankey89} and
Horsfield \cite{Horsfield97} ) used so far to calculate
exchange-correlation terms in these methods, using standard DFT
({\it e.g.} LDA). Then, in section III, we propose a new approach
that corrects the main deficiencies of previous approximations in
a practical manner (keeping always in mind computational
efficiency). In this approach, on-site (off-site)
exchange-correlation matrix elements are formally written as a
one-center (two-center) term plus a {\it correction} due to the
rest of the atoms. The one-center (two-center) term is evaluated
(and tabulated) directly, while the {\it correction} is calculated
using the SN approach. For this purpose, a general ({\it i.e.} for
arbitrary atomic-like basis set) version of the GSN approach has
been also developed.

The new scheme has been tested for some materials using the {\sc
Fireball} code and minimal $sp^3$ (for Al, Si and O) or $sp^3d^5$
(Au, Pd, W, Ti) basis sets. The results, presented in section IV,
show the good accuracy of the present first-principles TBMD
approach as compared with experiment and other accurate
calculations.

\begin{acknowledgments}
P.J. gratefully acknowledges financial support by the Spanish
Ministerio de Educacion, Cultura y Deportes.   This work has been
supported by the DGI-MCyT (Spain) under contracts MAT2001-0665 and
MAT2001-01534.
JPL and HW greatly acknowledges financial support from DOE Grant
No. DE-FG02-03ER46059 and from the Center for the Simulation of
Accidental Fires and Explosions (C-SAFE at the University of Utah),
funded by the Department of Energy, Lawrence Livermore National
Laboratory, under subcontract B341493.
OFS thanks the NSF (DMR 99- 86706) for support.
\end{acknowledgments}


{99}

\begin{thebibliography}{99}
\bibitem{Sankey89}
O.~F. Sankey and D.~J. Niklewski, Phys. Rev. B {\bf 40},  3979
(1989).
\bibitem{Hohenberg64}
P. Hohenberg and W. Kohn, Phys. Rev. {\bf 136},  B864  (1964).
\bibitem{Kohn65}
W. Kohn and L.~J. Sham, Phys. Rev. {\bf 140},  A1133  (1965).
\bibitem{Becke}
A.D.Becke, Phys. Rev. A {\bf 38}, 3098 (1988). 
\bibitem{Horsfield97}
A.P. Horsfield, Phys. Rev. B {\bf 56},  6594 (1997).
\bibitem{Horsfield00}
A.P. Horsfield and A.M. Bratkovsky, J. Phys.: Condens. Matter {\bf
12} R1 (2000).
\bibitem{Porezag} D. Porezag et al., Phys. Rev. B {\bf 51},
12947 (1995); T. Freuenheim et al. Phys. Rev. B {\bf 52}, 11492 (1995).
\bibitem{Siesta} P. Ordejon, E. Artacho, J.M. Soler, Phys. Rev. B {\bf 53},
10441 (1996); D. Sanchez-Portal, P. Ordejon, E. Artacho and J.M. Soler, Int.
J. Quant. Chem. {\bf 65}, 453  (1997).
\bibitem{OO-LCAO} P. Pou et al., Int. J. Quant. Chem.
{\bf 91}, 151 (2003); R. Oszwaldowski et al., J. Phys.: Cond. Matter {\bf 15},
S2665  (2003).
\bibitem{Ortega98a}
J. Ortega, Computational Materials Science {\bf 12},  192 (1998).
\bibitem{Eschrig78}
H. Eschrig and I. Bergert, Phys. Stat. Sol. B {\bf 90}, 621 (1978).
\bibitem{Junquera01}
J. Junquera, O. Paz, D. S\'anchez-Portal and E. Artacho, Phys.
Rev. B {\bf 64}, 235111 (2001).
\bibitem{Ortega00}
J. Ortega, R. P\'erez and F. Flores and A. Levy Yeyati, J. Phys.:
Condens. Matter {\bf 12}, L21 (2000).
\bibitem{Ortega98b}
J. Ortega, F. Flores and A. Levy Yeyati, Phys. Rev. B  {\bf 58},
4584 (1998).
\bibitem{Demkov95} A.A. Demkov, J. Ortega, O.F. Sankey and M.P. Grumbach,
 Phys. Rev. B {\bf 52}, 1618 (1995).
\bibitem{Lewis01}
J.P. Lewis {\it et al.}, Phys. Rev. B {\bf 64}, 195103 (2001).
\bibitem{Harris85}
J. Harris, Phys. Rev. B {\bf 31},  1770  (1985).
\bibitem{Foulkes89}
W. Foulkes and R. Haydock, Phys. Rev. B {\bf 39},  12520  (1989).
\bibitem{Lewis02}
J.P. Lewis, J. Pikus, Th. E. Cheatham, E.B. Starikov, H. Wang,
J. Tomfohr, and O.F. Sankey, Phys. Stat. Sol. {\bf 233}, 90 (2002).
\bibitem{FJ} 
F.J. Garcia-Vidal et al., Phys. Rev. B {\bf 50}, 10537 (1994).
\bibitem{Pou} 
P.Pou et al. Phys. Rev. B {\bf 62}, 4309 (2000).
\bibitem{exp-values} C. Kittel, Introduction to Solid State Physics, 6th ed. 
(Wiley, New York, 1986).
\bibitem{Kenny00}
S.D. Kenny, A.P. Horsfield and Hideaki Fujitani,
Phys. Rev. B {\bf 62}, 4899 (2000).
\bibitem{Anglada02}
E. Anglada, J.M. Soler, J. Junquera and E. Artacho,
Phys. Rev. B {\bf 66}, 205101 (2002).
\bibitem{Ozaki04}
T. Ozaki and H. Kino, Phys.
Rev. B {\bf 69}, 195113 (2004).
\bibitem{Papa}
D.A. Papaconstantopoulos, Handbook of the Band Structure of Elemental Solids,
(Plenum Press, New York, 1986). 

\bibitem{Burdett873639} J. K. Burdett, T. Hughbanks, G. J. Miller,
J. W. Richardson, and J. V. Smith, I. Am. Chem. Soc. {\bf 109},
3639(1987).
\bibitem{Chelikowsky921284}K. M. Glassford and J. R Chelikowsky,
Phys. Rev. B {\bf 46}, 1284 (1992).
\bibitem{Arlt0014414}T. Arlt, M. Bermejo, M. A. Blanco, L.
Gerward, J. Z. Jiang, J. S. Olsen, and J. M. Recio, Phys. Rev. B
{\bf 61}, 14414 (2000).
\bibitem{Mo9513023}S. D. Mo amd W. Y. Ching, Phys. Rev. B {\bf
51}, 13023 (1995).
\bibitem{Kowalczyk77161}S. P. Kowalczyk, F. R. McFeely, L Ley, V.
T. Gritsyna, and D. A. Shirley, Solid State Commun. {\bf 23}, 161
(1977).
\bibitem{Pascual785606}J. Pascual and H. Mathieu, Phys. Rev. B
{\bf 18}, 5606 (1978).
\bibitem{Vos773917}K. Vos, J. Phys. C: Solid State Phys. {\bf 10},
3917 (1977).
\bibitem{Kasowski795168}R. V. Kasowski and R. H. Tait, Phys. Rev.
B {\bf 20}, 5168 (1979).


\end{thebibliography}
\end{document}